%% file: main.tex
\documentclass[
    twocolumn,
	prd,
	amssymb,
	preprintnumbers,superscriptaddress,
	nofootinbib]{revtex4-1}

\pdfoutput=1

\usepackage{graphicx}
\usepackage{enumitem}
\usepackage{latexsym}
\usepackage{amsfonts}
\usepackage{amssymb}
\usepackage{xcolor}
\usepackage[export]{adjustbox}
\usepackage{amsmath}
\usepackage{slashed}
\usepackage{dcolumn}
\usepackage{verbatim}
\usepackage{float}
\usepackage{multirow}
\usepackage{xspace}
\usepackage[normalem]{ulem}
\usepackage[
pdfauthor={Djuna Croon}]{hyperref}

\input{universalnewcommands.tex}

\newcommand{\T}{{\rm T}}

\newcommand{\calO}{{\cal O}}

\allowdisplaybreaks

\begin{document}
\title{$\Delta B=2$ neutron decay into antiproton mode $n\to \bar pe^+\nu(\bar\nu)$}

\author{Xiao-Gang He}\email{hexg@phys.ntu.edu.tw}
\affiliation{Tsung-Dao Lee Institute, and School of Physics and Astronomy, Shanghai Jiao Tong university, Shanghai 20024, China}
\affiliation{Department of Physics, National Taiwan University, Taipei 106, Taiwan}
\affiliation{Physics Division, National Center for Theoretical Sciences, Hsinchu 300, Taiwan}

\author{Xiao-Dong Ma}\email{maxid@phys.ntu.edu.tw}
\affiliation{Department of Physics, National Taiwan University, Taipei 106, Taiwan}
\affiliation{School of Nuclear Science and Technology, Lanzhou University, Lanzhou 730000, China}

\preprint{}

\begin{abstract}
We discuss the unique baryon number violation by two units neutron decay mode $n\to \bar p e^+\chi$, with $\chi$ being the standard model  (SM) neutrino $\nu$ or antineutrino $\bar\nu$ or any beyond the SM light fermion, in the framework of effective field theory. This mode is kinematically allowed but rarely discussed theoretically or searched for experimentally. We estimate the lower bound on its partial lifetime from that of the dinucleon decay $np\to e^+\chi$ per oxygen nucleus $^{16}$O set by the Super-Kamiokande experiment, with a conservative bound $\Gamma^{-1}_{n\to\bar pe^+\chi}>5.7\times 10^{39}~\rm yrs$. We also discuss its characteristic signature for the future experimental search and astrophysical implications.
\end{abstract}

\maketitle

%\tableofcontents

%%%%%%%%%%%%%%%%%%%%%%%
\noindent 
{\bf Introduction.}
%%%%%%%%%%%%%%%%%%%%%%%
The baryon-antibaryon asymmetry of the universe requires the violation of baryon number ($B$), which is one of the three Sakharov conditions for a successful baryogenesis mechanism~\cite{Sakharov:1967dj}. The baryon number violation (BNV) is also a general feature in scenarios of physics beyond the standard model like the grand unified theories~\cite{Georgi:1974sy,Babu:1993we,Nath:2006ut}, in which the $|\Delta B|=1$ single nucleon decays like proton decay $p\to e^+\pi^0$ are predicted.  However, the scale of new physics (NP) associated with $|\Delta B|=1$ nucleon decay is constrained, by the current experimental data, to be around $10^{14-16}~\rm GeV$, which is unreachable to directly produce those heavy particles at current and future high energy colliders. 

On the other hand, there is a class of scenarios suppressing $|\Delta B|=1$ nucleon decay but contribute dominantly to $|\Delta B|=2$ processes like the neutron - antineutron ($n{\rm- }\bar n$) oscillation~\cite{Kuzmin:1970nx,Mohapatra:1980qe,Phillips:2014fgb}, hydrogen - antihydrogen (${\rm H}{\rm-}{\rm H}$) oscillation and dinucleon to dimeson/dilepton decays in nuclei (like $pp\to \pi^+\pi^+, e^+e^+$)~\cite{Nussinov:2001rb,Arnold:2012sd,Dev:2015uca, Gardner:2018azu,Girmohanta:2019fsx}.\footnote{A detailed summary on the $|\Delta B|=1,2$ physics can be found in 
Refs.~\cite{Heeck:2019kgr, Babu:2020nnh}.} In such scenarios, the associated NP scale is lowered to be around ${\cal O}(1-10^4)~\rm TeV$ from the current experimental results on $n{\rm- }\bar n$ oscillation~\cite{BaldoCeolin:1994jz,Abe:2020ywm} and dinucleon to dimeson/dilepton decays~\cite{Takhistov:2015fao,Takhistov:2016eqm,Sussman:2018ylo}, and this TeV scale NP is appealing since it may be directly tested at colliders like the LHC and other proposed ones through the search of the mode $pp\to \ell^+_1\ell^+_2+4{~\rm jets}$~\cite{Bramante:2014uda}. In this latter case with $|\Delta B|=2$ dinucleon decays, we noticed that there exists a unique $|\Delta B|=2$ single free or bound neutron decay mode $n\to \bar pe^+\chi$ with $\chi$ being the standard model (SM) neutrino $\nu$ or anti-neutrino $\bar \nu$ or any new light fermion beyond the minimal SM such as a light sterile neutrino,  but such a mode has not yet been considered both theoretically and experimentally.\footnote{A brief mention of $n\to \bar pe^+\nu_e$ related to operators causing $n{\rm-}\bar n$ oscillation is given in \cite{Berezhiani:2018xsx,Berezhiani:2018pcp}.}  
In this letter we will investigate this neutron decay mode in the framework of effective field theory and point out its distinct experimental signature relative to other dinucleon decays for the guidance of future experimental searches.  

From the perspective of standard model effective field theory (SMEFT), the leading order interactions contributing to $n\to \bar pe^+\chi$ depend on the type of $\chi$. For the $\chi$ to be the electron neutrino $\nu_e$,  $n\to \bar pe^+\nu_e$ can be generated, at leading order, by the same dimension-9 (dim-9) operators mediating the $n{\rm- }\bar n$ oscillation through an insertion of a SM charged weak current, which is similar to the neutron-antineutron conversion through the same $n{\rm- }\bar n$ oscillation operators did by Gardner and Yan~\cite{Rao:1983sd, Caswell:1982qs,Gardner:2017szu}. For the muon and tau neutrinos $\nu_{\mu,\tau}$, the leading order interactions for $n\to \bar pe^+\nu_{\mu,\tau}$ appear at
dim 13,\footnote{They are not generated at dim 12 in that the SMEFT has the property that the operator's dimension is even (odd) if $|\Delta B-\Delta L|/2$ is even (odd)~\cite{Kobach:2016ami}. For our case here, $|\Delta B|=2$ and $\Delta L=0$ means the dimension of relevant operator is odd.} this is because the SM has a good lepton flavor symmetry and therefore $n\to \bar pe^+\nu_{\mu,\tau}$ cannot be realized like the decay $n\to \bar pe^+\nu_e$ by dim-9 interactions. 

On the other hand, for the antineutrino case, 
the decay modes $n\to \bar pe^+\bar\nu_{e,\mu,\tau}$ also violate lepton number by two units ($|\Delta L|=2$) and their leading order interactions appear at dim 12. After sending the Higgs field to its vacuum expectation value and expanding the quark and lepton doublet, we obtain the relevant effective interactions mediating $n\to \bar pe^+\chi$ with $\chi\neq \nu_e$ in the so-called low energy effective field theory (LEFT) below the electroweak scale~\cite{Jenkins:2017jig,Liao:2020zyx,Li:2020tsi,Murphy:2020cly}, which consisting of two up-type and four down-type quarks and a charged lepton current, i.e., having the configuration $(uudddde\chi)$. For the decay $n\to \bar pe^+\nu_e$, it is a little bit complicated since it can be generated at leading order from the same dim-9 $n{\rm- }\bar n$ oscillation operators with the structure $(uudddd)$ by inserting a SM four-fermion vertex in one of the four down-type quark legs. 

\begin{figure}%[h!]
\centering
\includegraphics[width=8.5cm]{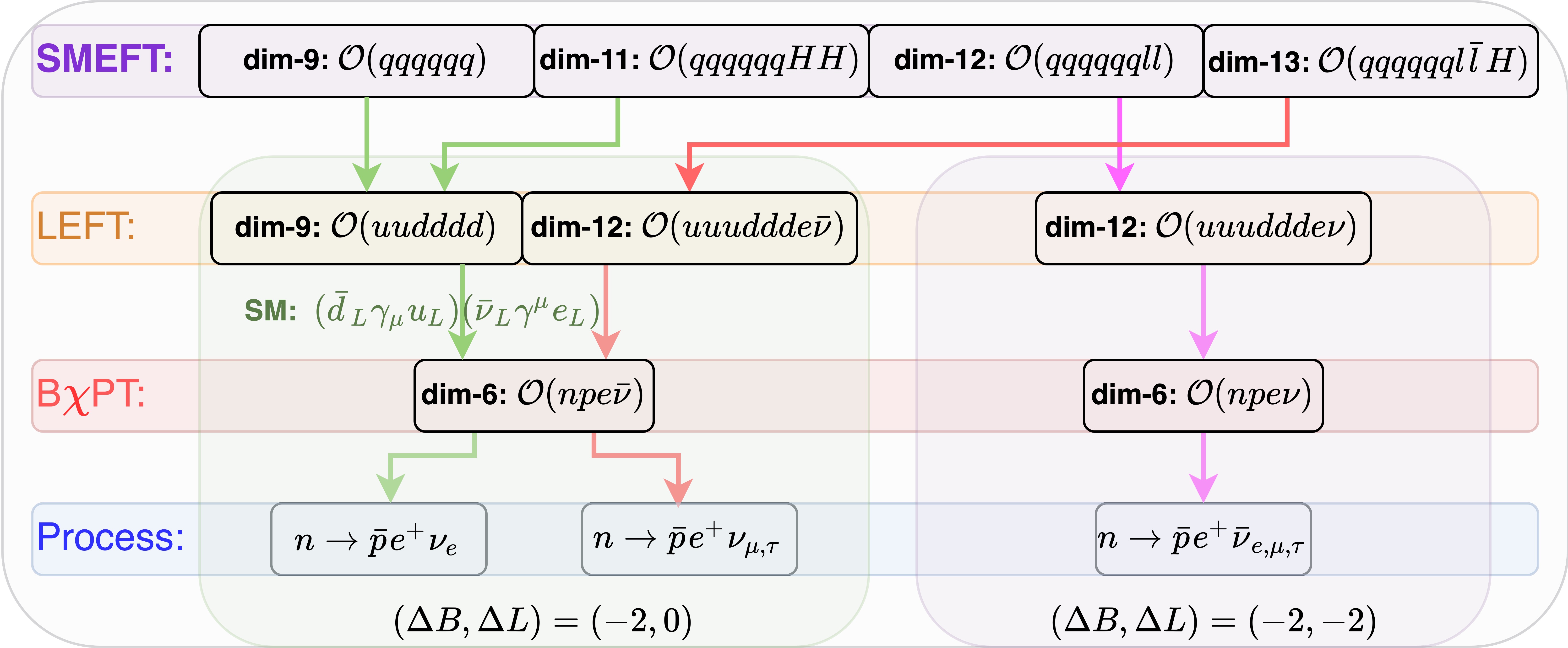}
\caption{A flowchart of EFT consideration of $n\to\bar p e^+\chi$.}
\label{figa}
\end{figure}

Once the relevant LEFT operators are obtained, one can perform a non-perturbative QCD matching for the six-quark sectors using the baryon chiral perturbation theory (B$\chi$PT) formalism~\cite{Jenkins:1990jv, Bernard:1995dp,Bijnens:2017xrz}. In this way, one ends up with operators consisting of nucleons, mesons and leptons and the decay rate can be readily calculated.  Here we note that the relevant interactions for $pn\to e^+\chi$ transition are the same as that of the decay $n\to\bar pe^+\chi$ because these two processes are related to each other by crossing symmetry. Therefore constraints on the branching ratio of $n\to\bar pe^+\chi$ can be obtained by using known bound from $pn\to e^+\chi$ transition. If realizing the interactions via SMEFT operators, these interactions may be related to other dinucleon to dilepton transitions such as $pp \to e^+e^+$ and $nn \to \bar \nu \bar \nu$. An example of such a case will be given below. We will present the details elsewhere for the above procedures to our accompanying long paper concerning the $|\Delta B=\Delta L|=2$ dinucleon to dilepton decays $(pp\to\ell^+\ell^{\prime+},pn\to\ell^+\bar\nu^{\prime},nn\to\bar\nu\bar\nu^{\prime} )$ in a full EFT analysis~\cite{He:2021mrt}. {In Fig.~\ref{figa}, we summarize our above discussion in a pictorial way. }
\\

%%%%%%%%%%%%%%%%%%%%%%%
\noindent
{\bf EFT analysis for $n\to\bar pe^+\chi$.}
%%%%%%%%%%%%%%%%%%%%%%%
In the following, we will take the EFT procedures outlined in the above and start directly from the hadronic level interactions for a model-independent analysis for $n\to\bar pe^+\chi$. We assume the dominant interactions contributing to $n\to\bar p e^+\chi$ are encoded in the leading order dim-6 operators with the field configuration $(npe\chi)$.\footnote{From the perspective of the SMEFT, the SMEFT operators generated above the electroweak scale should have all fields in contact form without additional propagators induced by SM degrees of freedom which could not lead to these dim-6 operators but rather even higher dimensional ones with additional derivatives in the form $(npe\chi\partial^n)$, we assume their effect relative to the dim-6 operators in Eqs.~(\ref{ope:npenubar},\ref{ope:npenu}) is small and can be neglected from the usual dimensional analysis.  } The relevant dim-6 effective $|\Delta B|=2$ interactions are classified into two sectors in terms of the net lepton number $|\Delta L|=0$ (for $\chi=\nu$) or $|\Delta L|=2$ (for $\chi=\bar\nu$) as follows
\begin{align}\nonumber
{\cal L}_{\Delta B=2}&=\sum_i \tilde C_{\Delta L=0,i}^{(pn)}\tilde  {\cal O}_{\Delta L=0,i}^{(pn)}
\\
&+ \sum_i C_{\Delta L=2,i}^{(pn)} {\cal O}_{\Delta L=2,i}^{(pn)}+{\rm h.c.}\;,
\label{effectiveL}
\end{align}
where all the independent dim-6 operators in each sector are parametrized as follows 
\begin{align}\nonumber
{n\to\bar pe^+\nu}:~%%%
&\tilde {\calO}^{(pn)S}_{R}=(p^\T Cn)(\overline{\nu_L}e_R)\;,
\\\nonumber
&\tilde {\calO}^{(pn)S}_{5R}=(p^\T C\gamma_5n)(\overline{\nu_L}e_R)\;,
\\\nonumber
&\tilde {\calO}^{(pn)V}_L=(p^\T C\gamma_\mu n)(\overline{\nu_L}\gamma^\mu e_L)\;,
\\\nonumber
&\tilde {\calO}^{(pn)V}_{5L}=(p^\T C\gamma_\mu\gamma_5 n)(\overline{\nu_L}\gamma^\mu e_L)\;,
\\
&\tilde {\calO}^{(pn)T}=(p^\T C\sigma_{\mu\nu}n)(\overline{\nu_L}\sigma^{\mu\nu}e_R)\;,
\label{ope:npenubar}
\\\nonumber
{n\to\bar pe^+\bar\nu}:~%%%
&{\calO}^{(pn)S}_{L}=(p^\T Cn)(e^\T_L C\nu_L)\;,
\\\nonumber
&{\calO}^{(pn)S}_{5L}=(p^\T C\gamma_5n)(e^\T_L C\nu_L)\;,
\\\nonumber
&{\calO}^{(pn)V}_L=(p^\T C\gamma_\mu n)(e^\T_R C\gamma^\mu\nu_L)\;,
\\\nonumber
&{\calO}^{(pn)V}_{5L}=(p^\T C\gamma_\mu\gamma_5 n)(e^\T_R C\gamma^\mu\nu_L)\;,
\\
&{\calO}^{(pn)T}=(p^\T C\sigma_{\mu\nu}n)(e^\T_L C\sigma_{\mu\nu}\nu_L)\;,
\label{ope:npenu}
\end{align}
where $C$ is the charge conjugation matrix satisfying $C^{\rm T}=C^\dagger=-C$ and $C^2=-1$, and the flavor of SM left-handed neutrino is suppressed. Except the neutron decay $n\to \bar p e^+\chi$, the above interactions can also contribute to the dinucleon decay $pn\to e^+\chi$ in nuclei and the conversion $e^-p\to \bar n\chi$ and $e^-n\to \bar p\chi$ in the electron-deuteron (e-d) scattering. This correlation of different processes due to the crossing symmetry can help us to estimate the partial lifetime of $n\to \bar p e^+\chi$ from the experimental lower bound on that of $pn\to e^+\chi$ per oxygen nucleus $^{16}$O set by the water Cherenkov Super-Kamiokande (SK) experiment~\cite{Takhistov:2015fao}. In the following, we assume each time there is only one operator dominant for the decays $n\to \bar p e^+\chi$ and $np\to e^+\chi$, then the bound on the latter mode can be used to extract a bound on the former. 

As mentioned earlier that the low energy effective interactions may be related to NP at high energies, which can be parametrized through the SMEFT interactions. Here we give an example how a SMEFT operator can generate some of the low energy operators. We find the dim-12 operator
\begin{align}\nonumber
{\cal O}_{Q^6L^2}^{S,(S)}&=(Q^{i\T}_a CQ^j_b)(Q^{k\T}_c CQ^l_d)(Q^{m\T}_e CQ^n_f)(L^\T_g CL_h^\prime)
\\
&\times\epsilon_{ab}\epsilon_{cd}\epsilon_{eg}\epsilon_{fh}T^{SAA}_{\{mn\}[kl][ij]}\;,
\end{align}
is of a very interesting case. Where $C$ is again the charge conjugation matrix, $Q$ and $L$ are the SM quark and lepton doublets, respectively. $T^{SAA}_{\{mn\}[ij][kl]}=\epsilon_{ijm}\epsilon_{kln}+\epsilon_{ijn}\epsilon_{klm}$ is a color tensor responsible for the operator to be color invariant, $\epsilon_{ab}(\epsilon_{ijk})$ is the second (third) rank totally antisymmetric Levi-Civita symbol. 

\begin{figure}%[h!]
\centering
\includegraphics[width=5cm]{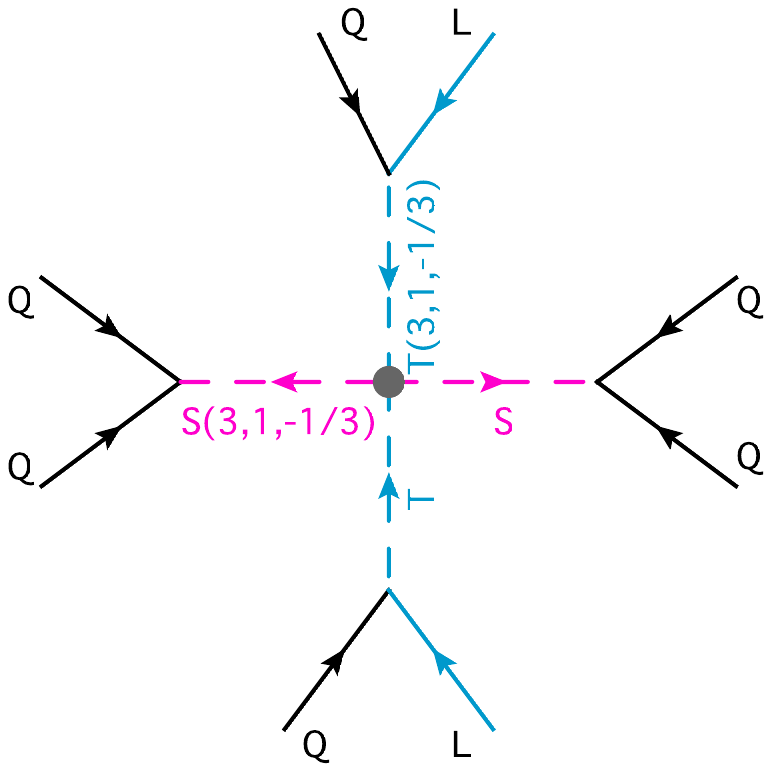}
\caption{The Feynman diagram responsible for the operator ${\cal O}_{Q^6L^2}^{S,(S)}$. Where the fields in blue are odd parity under $\mathbb{Z}_2$.}
\label{fig1}
\end{figure}

From UV completion viewpoint, this operator can be generated by introducing two leptoquark-like scalars $S(3,1,-1/3)$ and $T(3,1,-1/3)$ into the SM field spectrum and a $\mathbb{Z}_2$ symmetry, where the numbers indicate their SM quantum numbers under $SU(3)_C\times SU(2)_L\times U(1)_Y$. We arrange  the scalar $T$ and SM leptons $L, e$ are odd parity under $\mathbb{Z}_2$ and all others even. In this way, we have the Yukawa  interactions $(SQQ)$ and $(T^\dagger QL)$ and the Higgs quartic interaction $(S^\dagger T)^2$. Then one can easily see the Feynman diagram in Fig.~\ref{fig1} would yield the operator after integrating out those heavy scalars, as the same time, the $\mathbb{Z}_2$ symmetry forbids the generation of dim-6 $|\Delta B|=1$ operators like $(QQQL)$. In~\cite{He:2021mrt} we show there are only two independent operators with the form $(Q^6L^2)$, the other one is a `tensor-like' operator
\begin{align}\nonumber
{\cal O}_{Q^6L^2}^{T,(S)}&=(Q^{i\T}_a CQ^j_b)(Q^{k\T}_cCQ^l_d)(Q^{m\T}_eC\sigma_{\mu\nu}Q^n_f)\\
&\times(L^\T_g C\sigma^{\mu\nu}L_h^\prime)\epsilon_{ab}\epsilon_{cd}\epsilon_{ef}\epsilon_{gh}T^{SAA}_{\{mn\}[kl][ij]}\;,
\end{align}
we see from Fig.~\ref{fig1} this operator is also absent in our toy model. 
%%%%%%%%%%

The hadronic counterpart of the six-quark part of  ${\cal O}_{Q^6L^2}^{S,(S)}$ can be obtained through the B$\chi$PT matching, which has been used before for the dim-9 six-quark $n{\rm-}\bar n$ oscillation operators in~\cite{Bijnens:2017xrz}. We first identify that the six-quark part of ${\cal O}_{Q^6L^2}^{S,(S)}$ belongs to the irreducible representation $({\bf 3}_L, {\bf 1}_R)$ of the two-flavor QCD chiral group $SU(2)_L\times SU(2)_R$, then its leading order hadronic counterpart is realized by appealing to the nucleon field $\Psi=(p,n)^\T$ and pion matrix $u$ in such a way that the resultant operator has the same symmetry property as ${\cal O}_{Q^6L^2}^{S,(S)}$ including the Lorentz covariance, baryon number and chiral property.  After carrying out the procedures described in the above, we obtain the leading order hadronic operator taking the form $(u^\dagger)_{u_La}(u^\dagger)_{v_Lb} \Psi_{a}^\T C\left[ g_{3\times 1}+\hat g_{3\times 1} \gamma_5  \right]\Psi_b$ with $u_L, v_L=1,2$ for $u,d$ quarks.\footnote{One should be careful that this matching result can not be applied to processes involving hard pions like $pn\to e^+\chi+n\pi$ due to the breakdown of chiral power counting. In that case, one may employ the chiral effective theory for nuclear forces to reach a consistent theory, this has been recently applied to the study of deuteron decay from $n-\bar n$ oscillation operators~\cite{Dover:1985hk,Oosterhof:2019dlo,Haidenbauer:2019fyd}. }
Expanding to zeroth order in the pion fields,  we find the operator leads to ${\calO}^{(pn)S}_{L}$ and ${\calO}^{(pn)S}_{5L}$ in Eq.~\eqref{ope:npenu} with the coefficients 
\begin{eqnarray}
C_L^{(pn)S}=-8g_{3\times1}C_{Q^6L^2}^{S,(S)}\;, 
C_{5L}^{(pn)S}=-8\hat g_{3\times1}C_{Q^6L^2}^{S,(S)}\;,\;\;\;
\end{eqnarray}
where $C_{Q^6L^2}^{S,(S)}\equiv\Lambda_{\rm NP}^{-6}$ is the Wilson coefficient of operator ${\cal O}_{Q^6L^2}^{S,(S)}$ with $\Lambda_{\rm NP}$ being the associated NP scale. Here the hadronic low energy constants $g_{3\times1}$ and $\hat g_{3\times1}$ are pertinent to the non-perturbative QCD matching of quark level operators in the B$\chi$PT. $g_{3\times1}$ can be related to the $n{\rm-}\bar n$ oscillation matrix element via chiral symmetry, the latter has been determined by using lattice QCD method~\cite{Rinaldi:2019thf}, that implies $g_{3\times1}\sim 6\times 10^{-6}~\rm GeV^6$, while the value for $\hat g_{3\times1}$ has not been calculated. A naive dimensional analysis would give $\hat g_{3\times1}$ to be of order $\sim \Lambda_\chi^6/(4\pi)^4\sim 1.2\times10^{-4}~\rm GeV^6$~\cite{Weinberg:1989dx}, where $ \Lambda_\chi\sim 1.2~\rm GeV$ is the chiral symmetry breaking scale. This result can also be estimated from the fact that the dim-9 six-quark operator condensates into a dim-3 nucleon current together with the coupling $\hat g_{3\times1}$, the mass dimension of $\hat g_{3\times1}$ should be compensated by $\Lambda_{\rm QCD}\sim200~\rm MeV$ since it is the scale for the non-perturbative QCD effect, then $g_{3\times1}\sim \Lambda_{\rm QCD}^6\sim 6.4\times10^{-5}~\rm GeV^6$. We see both estimations agree within an $\calO(10)$ uncertainty. For numerical estimations, we will use this latter lower value for illustrations. The complete details for the above analysis will be presented in our accompanying long paper on LEFT and SMEFT operators~\cite{He:2021mrt}. 

It is also interesting that the above operator ${\cal O}_{Q^6L^2}^{S,(S)}$ not only generate $n\to \bar pe^+\bar\nu^\prime$ and $np\to \ell^+\bar\nu^\prime$, but also $pp\to\ell^+\ell^{\prime+}$ and $nn\to \bar\nu\bar\nu^\prime$. If the future SK experiments could find the $|\Delta B|=2$ signals from all those three different channels, it probably points to the NP incorporated in this operator.
\\

%%%%%%%%%%%%%%%%%%%%%%%
\noindent
{\bf Decay rate calculations}. 
%%%%%%%%%%%%%%%%%%%%%%%
From the above interactions in Eqs.~(\ref{effectiveL},\ref{ope:npenubar},\ref{ope:npenu}), it is straightforward to calculate the transition amplitude for $n(k)\to \bar p(p)e^+(p_1)\chi(p_2)$ and henceforth the decay rate. The decay rate can be written as
\begin{eqnarray}
\Gamma_{n\to\bar pe^+\chi}=\frac{1}{2m_n}\frac{1}{128\pi^3m_n^2}\int ds \int dt~\overline{\left|{\cal M}_{n\to \bar pe^+\chi}\right|}^2\;,\;\;
\label{decaywidth}
\end{eqnarray}
where $m_n$ is the neutron mass,  the Mandelstam variables are defined as $s=(p_1+p_2)^2,~t=(k-p_1)^2$ and $u=(k-p_2)^2$. 

Taking the interactions in Eqs.~(\ref{effectiveL},\ref{ope:npenubar},\ref{ope:npenu}) into consideration, after calculating the amplitude and finishing the phase space integration, and assuming one term dominates at a time, we obtain the following decay rate for each type of the couplings,
\begin{align}\nonumber
&\Gamma_{n\to \bar pe^+\nu}={m_n^5 \over 2^8\pi^3}\left\{ \epsilon_1\big|\tilde C_R^{(pn)S}\big|^2\;,\;\; \epsilon_2\big|\tilde C_{5R}^{(pn)S}\big|^2\;,\right. \\\nonumber
&~~~~~~~~~\left . \epsilon_3\big|\tilde  C_L^{(pn)V}\big|^2\;,\;\;\epsilon_4\big|\tilde C_{5L}^{(pn)V}\big|^2\;,\;\epsilon_5\big|\tilde C^{(pn)T}\big|^2 \right\}\;,
\\\nonumber
&\Gamma_{n\to \bar pe^+\bar\nu}={m_n^5 \over 2^8\pi^3}\left\{ \epsilon_1\big|C_L^{(pn)S}\big|^2\;,\;\;\epsilon_2\big|C_{5L}^{(pn)S}\big|^2\;, \right. 
\\
&~~~~~~~~~\left . \epsilon_3\big|C_L^{(pn)V}\big|^2\;,
\;\;\epsilon_4\big|C_{5L}^{(pn)V}\big|^2\;,\;\;\epsilon_5\big|C^{(pn)T}\big|^2 \right\}\;.
\label{decayrate1}
\end{align}
If several terms exist simultaneously, there are in general interference terms. In our numerical estimate later, we assume one term dominates at a time and therefore the interference terms are neglected. The dimensionless parameters $\epsilon_{i}$ stem from the phase space integration over the kinematic variables and take the following numerical values
\begin{align}\nonumber
&\epsilon_1=5.0\times 10^{-15}\;,
&&\epsilon_2=7.8\times 10^{-22}\;,
&&\epsilon_3=5.0\times 10^{-15}\;,
\\
&\epsilon_4=1.5\times 10^{-15}\;,
&&\epsilon_5=6.0\times 10^{-14}\;.
\end{align}
where the suppression of $\epsilon_{i}$ is because of the small phase space for the process. 

We now analyze the dinucleon to dilepton decay $pn\to e^+\chi$ in nucleus to obtain a correlation to the processes discussed above. From the same interactions in Eqs.~(\ref{effectiveL},\ref{ope:npenubar},\ref{ope:npenu}), we can calculate the transition rate for $pn\to e^+\chi$ in nucleus in the following manner~\cite{Goity:1994dq}
\begin{align}\nonumber
\Gamma_{p n\to e^+\chi}&={1\over (2\pi)^3\sqrt{\rho_p \rho_{n} }}\int d^3k_1d^3k_2\rho_p(k_1)\rho_{n}(k_2)
\\
&\times v_{\rm rel.}(1-\mathbf{v_1\cdot v_2})\sigma(pn\to e^+ \chi)\;,
\end{align}
where $\rho_p(k)(\rho_n(k))$ is the proton (neutron) density distribution in momentum space and $\rho_p (\rho_n)$ is the average proton (neutron) density. $\mathbf{v_1}(\mathbf{v_2})$ is the velocity of the nucleon $p(n)$ and $v_{\rm rel.}$ is their relative velocity $v_{\rm rel.}=|\mathbf{v_1}-\mathbf{v_2}|$. The total cross-section for the free nucleon scattering process $p(k_1)n(k_2)\to e^+(p_1)\chi(p_2)$ takes the form
\begin{align}
\sigma(pn\to e^+ \chi)={1\over 4E_1E_2v_{\rm rel.}}\int d\Pi_2\overline{\left|{\cal M}_{pn\to e^+ \chi}\right|}^{2},
\end{align}
where $E_{1}(E_{2})$ is the energy of the initial state nucleon $p (n)$.  $d\Pi_2$ is the relativistically invariant two-body phase space. For the oxygen nucleus $^{16}$O, we neglect the small effects due to 
nucleon Fermi motion and nuclear binding energy~\cite{Nishino:2012ipa}. Under the quasi-static approximation of nucleons, then the transition rate reduces into
\begin{align}
\Gamma_{pn\to e^+\chi}={\rho_N\over 4 m_N^2}\overline{\left|{\cal M}_{pn\to e^+\chi}\right|}^{2}\Pi_2\;,
\end{align}
where the average nuclear matter density $\rho_N$ approximately equals $0.25~{\rm fm}^{-3}$ for either proton or neutron, and $m_N=(m_n+m_p)/2$. The two-body final state phase space factor $\Pi_{2}=(1/ 8\pi)\left(1- {m_e^2 /(m_n+m_p)^2} \right)$. The relevant amplitude is calculated from the interactions in  Eqs.~(\ref{effectiveL},\ref{ope:npenubar},\ref{ope:npenu}). After finishing the phase space integration and combining all pieces together, we obtain
\begin{align}\nonumber
&\Gamma_{pn\to e^+\nu}={\rho_Nm_N^2\over 4\pi}
\left\{\eta_1 \big|\tilde C_{R}^{(pn)S}\big|^2\;,\;\;\eta_2 \big|\tilde C_{5R}^{(pn)S}\big|^2\;, \right .\\ \nonumber
&~~~~~~~\left . \eta_3\big|\tilde C_{L}^{(pn)V}\big|^2\;,\;\;\eta_4\big|\tilde C_{5L}^{(pn)V}\big|^2
\;,\;\;\eta_5\big|\tilde C^{(pn)T}\big|^2\right\}\;,\\\nonumber
&\Gamma_{pn\to e^+\bar\nu}={\rho_Nm_N^2\over 4\pi}
\left\{\eta_1 \big|C_{L}^{(pn)S}\big|^2\;,\;\;\eta_2 \big|C_{5L}^{(pn)S}\big|^2\;,\right .\\
&~~~~~~~\left . \eta_3\big|C_{L}^{(pn)V}\big|^2\;,\;\;\eta_4\big|C_{5L}^{(pn)V}\big|^2
\;,\;\;\eta_5\big|C^{(pn)T}\big|^2\right\}\;,
\label{decayrate2}
\end{align}
where again we neglect the interference terms between any pair of Wilson coefficients under the assumption of one operator dominant each time. $\eta_i$ are dimensionless parameters and defined as 
\begin{align}\nonumber
&\eta_1=0\;, 
\eta_2=\left(1-{\delta^2 \over 4}\right)^2=1\;,
\\\nonumber
&\eta_3={\delta^2\over4}\left(1-{\delta^2 \over 4}\right)^2=7.4\times 10^{-8}\;,
\\\nonumber
&\eta_4=\left(2+{\delta^2\over4}\right)\left(1-{\delta^2 \over 4}\right)^2=2\;,
\\
&\eta_5=4\left(1+{\delta^2\over2}\right)\left(1-{\delta^2 \over 4}\right)^2=4\;,
\end{align}
where $\delta=m_e^2/m_N^2$ and the nucleon velocity effect is neglected. If we keep the small  nucleon velocity effect,  $\eta_{1,3}$ will be modified to be proportional to the squared velocity, i.e., $\tilde \eta_{1,3}\sim v_N^2\sim{2\Delta E/m_N}\sim 10^{-2}$. The nucleon velocity $v_N$ is estimated to be around $0.1~c$ by taking the average nucleon binding energy $\Delta E\sim 8\rm~MeV$ in oxygen nucleus, where $c$ is the speed of light. For $\eta_{2,4,5}$, the nucleon velocity effect is small relative to their leading contribution and can be safely neglected. In the following, we use the modified value $\tilde \eta_{1}\sim v_N^2\sim 10^{-2}$ for $\eta_1$ but still keep $\eta_{3}$'s value as a conservative estimation for the bound on partial lifetime of $n\to \bar pe^+\chi$. 

In order to employ the experimental result for $np\to e^+\chi$, we assume each time there is only one operator dominant. According to the decay rate in Eq.~\eqref{decayrate1} for $n\to \bar pe^+\chi$ and Eq.~\eqref{decayrate2} for $np\to e^+\chi$, after keeping one term each time and eliminating the common Wilson coefficient, we obtain the following relation
\begin{align}\nonumber
\Gamma_{n\to \bar pe^+\chi}^{-1}
&={2^8\pi^3\over m_n^5}{\rho_Nm_N^2\over 4\pi}{\eta_i \over \epsilon_i}\Gamma_{pn\to e^+\chi}^{-1}
\\\nonumber
&={2^6\pi^2\rho_N\over m_n^3}{\eta_i \over \epsilon_i}\Gamma_{pn\to e^+\chi}^{-1}
\\
&\gtrsim {2^6\pi^2\rho_N\over m_n^3}{\eta_i \over \epsilon_i}\Gamma_{pn\to e^+\chi}^{-1,\rm SK}
\;,
\end{align}
where in the last step we have taken the SK lower bound on the partial lifetime of $pn\to e^+ \chi$ as our input, in which $\Gamma_{pn\to e^+\chi}^{-1,\rm SK} \gtrsim 2.6\times 10^{32}~\rm yrs$~\cite{Takhistov:2015fao}. Then our main result for the prediction of the $|\Delta B|=2$ neutron decay $n\to \bar pe^+\chi$ is as follows
\begin{align}
&\Gamma_{n\to \bar pe^+\chi}^{-1} \gtrsim 7.7\times 10^{44}~\rm yrs\;,
&&{\rm for}~\tilde{\cal O}_R^{(pn)S}\;, {\cal O}_L^{(pn)S}\;,
\\
&\Gamma_{n\to \bar pe^+\chi}^{-1} \gtrsim 4.9\times 10^{53}~\rm yrs\;,
&&{\rm for}~ \tilde{\cal O}_{5R}^{(pn)S}\;,{\cal O}_{5L}^{(pn)S}\;,
\\
&\Gamma_{n\to \bar pe^+\chi}^{-1} \gtrsim 5.7\times 10^{39}~\rm yrs\;,
&&{\rm for}~\tilde{\cal O}_L^{(pn)V}\;,{\cal O}_L^{(pn)V}\;,
\\
&\Gamma_{n\to \bar pe^+\chi}^{-1} \gtrsim 5.2\times 10^{46}~\rm yrs\;,
&&{\rm for}~\tilde{\cal O}_{5L}^{(pn)V}\;,{\cal O}_{5L}^{(pn)V}\;,
\\
&\Gamma_{n\to \bar pe^+\chi}^{-1} \gtrsim 2.6\times 10^{46}~\rm yrs\;,
&&{\rm for}~\tilde{\cal O}_R^{(pn)T}\;,{\cal O}^{(pn)T}\;.
\end{align}
We see that the extrapolated lower bound on the partial lifetime varies from ${\cal O}(10^{39})$ to ${\cal O}(10^{53})$ years depending on which operator is dominant. Furthermore, if we take $\tilde \eta_3\sim 10^{-2}$ as did for $\tilde \eta_1$, the lower bound is improved to be ${\cal O}(10^{45})$ years.  As a comparison with the experimental bound on the inclusive mode $\Gamma_{n\to e^+\rm anything}^{-1,\rm PDG}\gtrsim 0.6\times 10^{30}~\rm yrs$ quoted by the Particle Data Group~\cite{Zyla:2020zbs}, we find the indirect bound obtained here are improved by at least ten orders of magnitude. For the $\chi=\nu_e$ case, Ref.~\cite{Girmohanta:2019cjm} obtained a strong bound on $\Gamma_{pn\to e^+\nu_e}^{-1}\gtrsim10^{41}~\rm yrs$, which shall shift our above lower bound on $n\to\bar pe^+\nu_e$ by nine orders of magnitude or so.

Even though we obtain the above bound on the SM neutrino case, the bound is also valid for  beyond the SM light fermion with a mass $m_\chi<m_n-m_p-m_e$, the analysis is similar to the above SM neutrino case. If the mass $m_\chi\neq0$, we would expect a stronger bound on the partial lifetime due to an even smaller phase space.

We can also use the known constraints on dinucleon to dilepton transition rate to set bound on NP scale. Applying the SK experimental limits on the transitions $(pp\to\ell^+\ell^{\prime+},pn\to\ell^+\bar\nu^{\prime},nn\to\bar\nu\bar\nu^{\prime} )$ from our example operator ${\cal O}_{Q^6L^2}^{S,(S)}$~\cite{Takhistov:2015fao,Sussman:2018ylo}, we find the scale of NP $\Lambda_{\rm NP}\gtrsim 2-3~{\rm TeV}$~\cite{He:2021mrt}, which is compatible with the analysis from collider study in~\cite{Bramante:2014uda}. In~\cite{He:2021mrt}, we also set constraints on the Wilson coefficients of the dim-6 hadronic operators in Eq.~\eqref{ope:npenu} and the new physics scale associated with the relevant dim-12 SMEFT operators by using the procedures outlined above together with the experimental limits on the dinucleon decays.\footnote{We also noticed Ref.~\cite{Babu:2003qh} has considered the $|\Delta B|=3$ triple nucleon decay processes from dim-15 SMEFT operators, the inferred NP scale is around ${\cal O}(100~\rm GeV)$.} This may be probed at the LHC by looking at the process $pp\to \ell^+\ell^{\prime+}+4{\rm~jets}$ or the future LHeC via $e^-p\to \ell^++5{\rm~jets}$.  
\\

%%%%%%%%%%%%%%%%%%%%%%%
\noindent
{\bf Discussions.}
%%%%%%%%%%%%%%%%%%%%%%%
From our analysis in this paper, we expect the partial lifetime for the unique $|\Delta B|=2$ neutron decay mode $n\to \bar pe^+\chi$ is far longer than the current experimental sensitivity. Nevertheless, this decay mode has several distinct features in experimental searches and also strong astrophysical implications.  First, in the bound nuclei, if a neutron decays into an antiproton and a positron, then the sub-MeV positron can be detected. The antiproton can annihilate with a neighbour proton(neutron) to release two bunch of energetic mesons in the opposite direction through the QCD interactions, or two ${\cal O}(1~\rm GeV)$ gamma photons via the QED interaction, those correlated signals can be used  to identify this decay mode if it could happen. It is worth of search in the future neutrino experiments like the Super-K, DUNE~\cite{Abi:2020kei}, JUNO, etc. Second, unlike the $|\Delta B|=2$ dinucleon to dimeson/dilepton decays~\cite{Takhistov:2015fao,Takhistov:2016eqm,Sussman:2018ylo}, this decay mode can also be looked for from free neutron decay and oscillation experiments like the European Spallation Source~\cite{Addazi:2020nlz}  to obtain direct limit to test possibilities in all different ways despite the strong indirect bound given in this work. This neutron decay mode could also have impact on the astrophysical processes like the cooling of neutron star and the evolution of the universe, which can be used to constrain the relevant NP complementarily.

%%%%%%%%%%%%%%%%%%%%%%%
\section*{Acknowledgement}
%%%%%%%%%%%%%%%%%%%%%%%

This work was supported in part by NSFC (Grants 11735010, 11975149, 12090064), by Key
Laboratory for Particle Physics, Astrophysics and Cosmology, Ministry of Education, and Shanghai
Key Laboratory for Particle Physics and Cosmology (Grant No. 15DZ2272100), and in part
by the MOST (Grants No.109-2112-M-002-017-MY3 and 109-2811-M-002-535). XDM would like to thank Fu-Sheng Yu for his invitation as a visitor at Lanzhou Uni. due to the influence of the COVID-19 on traveling back to Taiwan, and also the valuable discussions with him on this work.

%%%%%%%%%%%%%%%%%%%%%%%
%%%%%%%%%%%%%%%%%%%%%%%
\bibliography{references}

\end{document}

%% file: universalnewcommands.tex
\renewcommand{\tilde}{\widetilde} 
\newcommand{\beq}{\begin{equation}}
\newcommand{\eeq}{\end{equation}}
\newcommand{\bea}{\begin{eqnarray}}
\newcommand{\eea}{\end{eqnarray}}